\documentstyle[12pt,epsf]{article}
\newcommand{\lsim}{\mbox{\raisebox{-.3em}{$\stackrel{<}{\sim}$}}}

\renewcommand{\cite}[1]{\ref{#1}}

\newcommand{\half}{\frac{1}{2}}
\newcommand{\beq}{\begin{equation}}
\newcommand{\eeq}{\end{equation}}
\newcommand{\beqa}{\begin{eqnarray}}
\newcommand{\eeqa}{\end{eqnarray}}
\newcommand{\bcent}{\begin{center}}
\newcommand{\ecent}{\end{center}}

\begin{document}
\baselineskip=0.6cm
\begin{center}
{\Large\bf NONZERO $\Omega_{\Lambda}$ AND A NEW TYPE OF\vspace{.2em}\\
THE DISSIPATIVE STRUCTURE\footnote{Based on the talk delivered at XXXIIIrd Rencontres de
Moriond, Fundamental Parameters in Cosmology (Les Arcs, France,
January 17-24, 1998)}}\vspace{.6em}\\
Yasunori Fujii\footnote{E-mail address: fujii@handy.n-fukushi.ac.jp}\vspace{.6em}\\
Nihon Fukushi University, Handa, 475-0012\ Japan\\
and\\
ICRR, University of Tokyo, Tanashi, Tokyo, 188-8502\ Japan
\end{center}
\baselineskip=0.6cm
\bigskip
\bcent
{\large\bf Abstract}\\
\bigskip
\begin{minipage}{13cm}
We revisit the proposed theoretical model for a small but nonzero
cosmological constant which seems supported increasingly better by
recent observations.  The model  features two scalar fields
which interact with each other through a specifically chosen nonlinear
potential.  We find a very sensitive dependence of the solutions of
the scalar field equations on the initial values.  We discuss how the
behavior is similar to and different from those in well-known chaotic
 systems, coming to suggest an interesting
new type of the dissipative structure.
\end{minipage}

\ecent

The cosmological constant problem has two faces; an upper bound and a
lower bound.  The former is given by the critical density $\rho_{\rm cr}$, or $\Omega_{\Lambda}\equiv \Lambda /\rho_{\rm cr} \lsim 1$.  In the Planckian unit system with $ 8\pi G =c= \hbar =1$, we know $\rho_{\rm cr}\equiv
3 H^2_0 \sim t_0^{-2}$, and the present age of the universe $t_0 \sim 10$Gy is of the order of $10^{60}$; hence the upper bound $\Lambda_{\rm ob} \lsim 10^{-120}$.  On the other hand, almost any theoretical models of unification gives $\Lambda_{\rm th} \sim 1$. The discrepancy  is the well-known disaster.   Recent observations seem to indicate  a lower bound as well;  $\Omega_{\Lambda}\sim 0.7$.   Let us try to understand this  ``small but nonzero cosmological constant."

A solution for the upper bound may come from  a scalar field $\phi$ of the Brans-Dicke type:
\beq
{\cal L}=\sqrt{-g}\left( \half\xi\phi^2 R -\half\epsilon g^{\mu\nu}\partial_{\mu}\phi \partial_{\nu}\phi +\Lambda +L_{\rm matter}
\right),
\label{chs1_3}
\eeq
where $\xi$ is a dimensionless constant
related to BD's parameter $\omega$ by $\omega\xi =1/4$, and $\epsilon
=\pm 1$.  The scalar field contribution is shown to cancel the effect of $\Lambda$ asymptotically.  To see  this, it is most convenient to apply a conformal transformation moving to the Einstein frame (as denoted by $*$) with nonminimal coupling removed.  The same Lagrangian as  Eq.~\ref{chs1_3} is now put into
\beq
{\cal L}=\sqrt{-g_*}\left( \half R_* -\half g^{\mu\nu}_* \partial_{\mu}\sigma \partial_{\nu}\sigma -V(\sigma) +L_{*\rm matter} \right),
\label{chs1_4}
\eeq
where the new canonical scalar field $\sigma$ is related to the original $\phi$; $\phi =\xi^{-1/2}e^{\zeta\sigma}$, with $\zeta =(6 + \epsilon \xi^{-1})^{-1/2}$.  We also find that $\Lambda$ now acts as a potential $V(\sigma)$:
\beq
V(\sigma)=\Lambda e^{-4\zeta\sigma},
\label{chs1_6}
\eeq
which pushes $\sigma$  toward infinity; the effect of $\Lambda$ continues to decrease.

In spatially flat RW cosmology, the cosmological equations are
\beq
3H^2 =\rho_{\sigma} +\rho_{\rm m} \equiv \rho,
\quad\mbox{and}\quad
\ddot{\sigma}+3H\dot{\sigma} + V'(\sigma)=0,
\label{chs1_7} 
\eeq
where we omit the symbol $*$ to simplify the notation.  We should also supply the equation for the matter density $\rho_{\rm m}$ as usual. Notice that $\rho_{\sigma}=\half \dot{\sigma}^2 +V(\sigma)$ is the energy density of the scalar field to be interpreted as an ``effective cosmological constant" $\Lambda_{\rm eff}$.

As an asymptotic solution, we obtain (for assumed radiation dominance)
\beqa
a(t)&=& t^{1/2}, \quad\mbox{and}\quad \sigma (t)= \bar{\sigma}+\half \zeta^{-1}\ln t,
\label{chs1_11}\\
\rho(t)&=& \frac{3}{4}\left( 1-\frac{1}{4}\zeta^{-2} \right) t^{-2},
\quad\mbox{with}\quad  \rho_{\sigma}(t)=\frac{3}{16}\zeta^{-2} t^{-2}.\label{chs1_13}
\eeqa
The last equation shows that $\Lambda_{\rm eff}=\rho_{\sigma}$ does
decay like $t^{-2}$, hence implementing the scenario of a ``decaying
cosmological constant."  {\em Today's $\Lambda$ is small only because
our universe is old} [\cite{fn1}].

We notice, however, that $\Lambda_{\rm eff}(t)$ falls off in the same way as the ordinary matter density $\rho_{\rm matter}(t)$ does.  Not qualified to be called a ``constant,"  it is simply
another form of dark matter.  It does not help understanding the lower bound.  We must include something that behaves more or less like a constant or at least falls off more slowly than $t^{-2}$.  On the other hand, the behavior $\sim t^{-2}$ was the key to have a ``small" $\Lambda$.  As a compromise we expect that $\Lambda_{\rm eff}(t)$ would decay like $\sim t^{-2}$ as an overall behavior, but with some  ``landings,"  in one of which  we today happen to be, and which  mimics a cosmological ``constant" for some duration of time.

Trying to implement the idea, we came across a model on a
try-and-error-basis, by introducing another scalar field $\Phi$ which
couples  to $\sigma$ [\cite{fn2},\cite{yf}].   The  Lagrangian is the same as Eq.~\ref{chs1_4} with the added kinetic term of $\Phi$ and $V(\sigma)$ replaced by (see Fig. 2 of Ref. 2)
\beq
V(\sigma, \Phi)= e^{-4\zeta \sigma}\left(
\Lambda +\half m^2 \Phi^2 \left[ 1+\beta \sin (\kappa \sigma) \right]\right),
\label{chs1_16}
\eeq
where $m, \beta, \kappa$ are the constants naturally of the order unity.  It has a ``central valley" given by Eq.~\ref{chs1_6}, but shows an  oscillatory behavior $\sim \sin (\kappa\sigma)$ if we climb the wall in the direction of $\Phi$.

\begin{figure}[tbh]

\epsfysize=13cm
\hspace*{2.5cm}
\epsffile{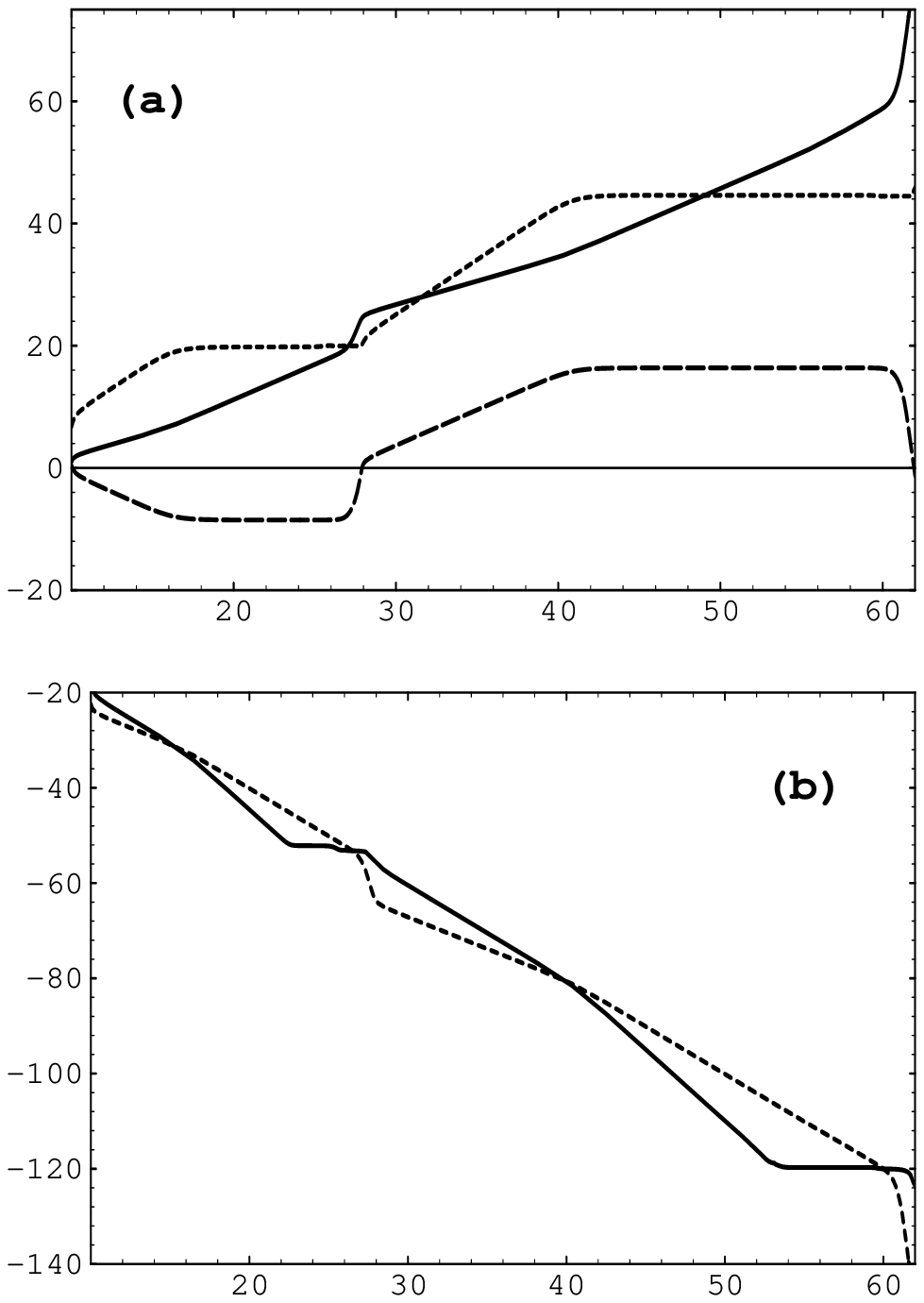}
\caption{(a)\ $\ln a$ (solid; $a$ for the scale factor), $\sigma$
(dotted), $2\Phi$ (broken) against $\log t$.  (b)\ $\log\rho_s$ (solid), $\log\rho_{\rm m}$ (dotted).  The parameters are $\Lambda =1, \zeta=1.582, m=4.75, \beta =0.8, \kappa =10.$  The initial values at $t_1 =10^{10}$ are $\sigma_1 =6.75442, \dot{\sigma}_1 =0, \Phi_1 =0.212, \dot{\Phi}_1=0$.  At $\log t_0 = 60.15$ corresponding to the present age chosen to be $1.21\times 10^{10}$y, we obtain $H_0 =81{\rm km}/{\rm s}/{\rm Mpc}$ and $\Omega_{\Lambda}=0.67$. See also Captions to Figs. 1 and 2 in Ref. 3.\hspace{\fill}
 }
\label{fg:1}
\end{figure}

A typical solution is shown in Fig.~\ref{fg:1}.  We started the integration at some time in the post-inflation era.  We also {\em mildly} fine-tuned initial conditions such that $\Omega_{\Lambda}$ somewhat smaller than 1, say, at $t_0 \approx 10^{60}$.  Notice a sequence of alternate occurrence of a rapid change and a nearly standstill of the  scalar fields.  Also toward the end of each ``landing" of $\rho_s(=\Lambda_{\rm eff}(t))$, which is now the energy density of the coupled $\sigma$-$\Phi$ system, the scale factor $a(t)$ shows a mini-inflation, an expected behavior.  Magnifying a portion of Fig.~\ref{fg:1}(b) around the present time $(\log t_0 \sim 60)$ extending to a tiny interval $(\Delta\log t \sim 1)$ yields a ``practical" plot going back to $z\sim 5$, showing $\Lambda_{\rm eff}$ which looks nearly constant.

These landings, as indicated also by other examples,  tend to occur {\em nearly periodically} with respect to $\ln t$.  This seems to suggest that some cosmological anomalies might have taken place nearly periodically with respect to $\ln t$.

As we find, the results depend on the choice of the parameters quite sensitively, particularly on the initial conditions.  In this connection, we first point out that the most important ingredients are present in the dynamics of the system of the two scalar fields.  We find the alternate occurrence of the ``wake-up phase" and the ``dormant phase" of the scalar fields even without the cosmological environments, provided the frictional coefficients $3H$ vary as $\sim t^{-1}$.

\begin{figure}[tbh]

\epsfysize=7.5cm
\vspace{-2.5cm}
\hspace*{2.5cm}
\epsffile{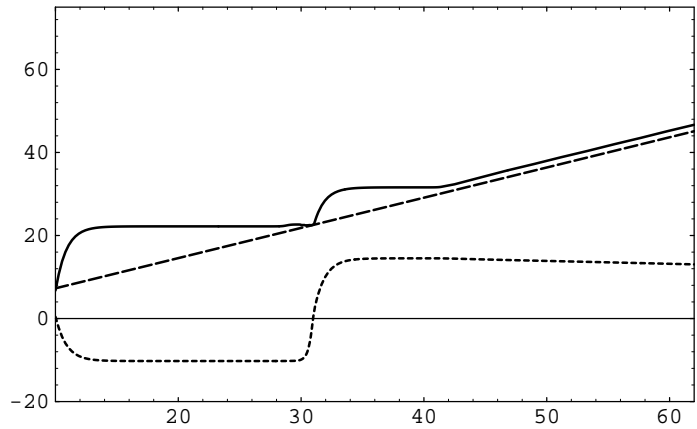}
\vspace{.5cm}
\caption{An example of the two-scalar system, in which the solution enters the asymptotic behavior after two repeated patterns. $\sigma$ (solid) and $2\Phi$ (dotted) and the broken line for the asymptotic behavior of $\sigma$. $\Phi$ will tend  to zero slowly.  The parameters and the initial values are the same as those in Fig. 1 except for the frictional coefficient chosen to be $(3/2)t^{-1}$.  See also Caption to Fig. 3 in Ref. 3.  A slight change of the initial values may result in a sudden change of the repetition number to three, for example, also changing the time of the end of the repetition period.\hspace{\fill} }
\label{fg:3}
\end{figure}

Fig.~\ref{fg:3} is one of the examples of the solutions of the isolated 2-scalar system with the same potential with the frictional constants $\Gamma =\gamma /t$.  Suppose we start from an initial position off the central valley on the potential Eq.~\ref{chs1_16}.  $\sigma$ will be pushed in the positive direction.  The potential will decrease quickly because it falls off like $e^{-4 \zeta \sigma}$.  Soon both fields will be virtually free.  Due to the frictional forces they will be decelerated almost to a complete stop.  However, as the time goes on, the frictional forces will also decrease according to $\sim t^{-1}$.  They will become even weaker than the forces derived from the potential which had  much dwindled.  Then the fields are suddenly pushed again.  But the potential will be weakened soon again
leading to the next dormant period.

However, in order for the fields to  be pushed  strongly after the dormant phase, their positions and velocities must ``match" the previous values to some extent.  Otherwise, the fields will not be waken up, beginning  simply to fall the potential slope slowly toward and along the central valley.  This is  the asymptotic behavior,  as we find in this example, reached commonly for any initial conditions.  Obviously this corresponds to a ``fixed-point attractor," like in standard damped oscillators.

The ``matching condition" is rather subtle.  This is the reason why a
slight change of initial values may sometimes result in a considerable change in the behaviors afterward.   This  sensitivity to the initial values reminds us of a well-known {\em chaotic behavior in nonlinear
systems.}  Sudden change of the number of  repetitions before reaching the smooth asymptotic behavior, for example, might be compared to sudden occurrence of ``bifurcation" in typical chaotic solutions.

However, there is an important difference from purely chaotic systems.  Usually a chaotic behavior is characterized by ``strange attractors," while our solutions tend to a simple fixed-point attractor.  Complicated hard-to-predict states occur only during a finite transient period.  In accordance with this no fractal structure is produced in the orbits in  phase space;  fractal structure might be present only to a limited ``depth."

We thus find what might be called   an ``incomplete chaos" or an ``incomplete fractals.'' They can be still interesting because genuine chaos might be a mathematical idealization for natural phenomena.  Also  chaotic behaviors are seen
almost every aspect of life.  But obviously life lasts only during a
limited lifetime; the ultimate destination  is a
non-chaotic death, corresponding to a fixed-point attractor.  In this sense, our model might provide  a primitive example of a living system.  On the other hand, it seems to  give a model of pattern formation (with variable period).

In our model, near periodicity in $\ln t$ is a consequence of $\Gamma
\sim t^{-1}$.  There are many examples of pattern formation with some
periodicity. Patterns on sea shells, for example, can be reproduced by
certain nonlinear equations of rather simple structure
[\cite{shells}].  Periods of these patterns are not determined by the
restoring force but, unlike in simple systems, primarily by friction
or dissipation in addition to the coupling strength.  This feature is shared in our model.   In this sense
also, our nonlinear equations might provide another example of
``dissipative structure."  It is interesting to notice that a model of
game dynamics with more than three species also generates periodicity
with respect to $\ln t$ [\cite{chw}].

Did we introduce as many degrees of freedom and parameters as we need?  The answer: the cosmological constant problem is so challenging that we may not have reached even close to any promising result no matter how many
ingredients  we introduced if we have come in a wrong direction.  It
seems  we are on a right track.

We thank Kunihiko Kaneko and Takashi Ikegami for enlightening discussions
 on nonlinear systems.

\bigskip
\bcent
{\large\bf References}
\ecent

\begin{enumerate}
\item\label{fn1}Y. Fujii and T. Nishioka, Phys. Rev. {\bf D42}(1990),
361, and papers cited therein.
\item\label{fn2}Y. Fujii and T. Nishioka, Phys. Lett. {\bf 254}(1991), 
347.
\item\label{yf} Y. Fujii, Astrop. Phys. {\bf 5}(1996), 133.
\item\label{shells}H. Meinhardt, {\sl The algorithmic beauty of sea
shells}, Springer-Verlag, 1995.
\item\label{chw}T. Chawanya, Prog. Theor. Phys. {\bf 94}(1995), 163.
\end{enumerate}

\newpage
\vspace*{2.5cm}

\newpage
\vspace*{4.5cm}

\end{document}